\documentstyle[prl,aps,multicol,epsf]{revtex}
\begin{document}
\title{Spin-only approach to quantum magnetism in the ordered stripe phase.}
\author{J. Tworzyd\l o$^a$, C. N. A. van Duin and J. Zaanen}
\address{Lorentz Institute, Leiden University, P.O.B. 9506, 2300 RA Leiden,
The Netherlands\\
$^a$ on leave from Institute of Theoretical Physics,
Warsaw University}
\date{\today}
\maketitle

\begin{abstract}
It has been argued that the spin-dynamics in cuprate superconductors is 
governed by the proximity to a zero-temperature critical point. This 
critical point would be
related to a transition from the superconducting phase to an ordered stripe state.
Using a coupled spin-ladder model, we investigate to what extent the strong
quantum spin-fluctuations in the ordered stripe state can be attributed to the spin sector. 
For the
bond-centered stripes, we find that our spin-only model can account
for the observed spin-fluctuations. For site-ordered stripes,
coupling of the spin and charge sector is needed.
\end{abstract}
\pacs{64.60.-i, 71.27.+a, 74.72.-h, 75.10.-b}

\begin{multicols}{2}
\narrowtext
Recent experiments on magnetism in the underdoped cuprates indicate
that the spin-sector in these materials is close to a zero-temperature
critical point \cite{BP,Aepli}. It was suggested by Aepli {\em et. al.}
that this critical point is due to
a transition from the spin-disordered superconducting phase, to a spin-ordered stripe phase \cite{Aepli}. To reach this, an
extra parameter has to been tuned, apart from hole-doping. This can be
for instance LTT deformation, which stabilizes stripe-ordering, or 
Zn-doping/magnetic field, which suppresses superconductivity.

The transition seems to be in the universality class of the Quantum Non-Linear Sigma model (QNLS) \cite{BP,Aepli,SC}. What the properties of the QNLS are
and how they fit the observed behavior will be discussed below. First, we 
should point out that it is actually not self-evident that this model
should capture the long-wavelength magnetic behavior in the superconducting
phase. The QNLS is a model of transversal spin fluctuations that has been
successfully applied to the antiferromagnetic phase at zero doping \cite{CHN}.
Since the microscopy of the superconducting phase is very different from that
of the zero-doping antiferromagnet, it would be remarkable if both
systems showed the same magnetic behavior at long distances.

In this paper, we will focus on the stripe phase, at
the spin-ordered side of the transition. This phase provides a perspective on how to arrive at the critical behavior, starting from the microscopy. Consider 
the following cartoon of a charge ordered stripe state. 
It consists of a regular grid of
lines of immobile holes, which form anti-phase domain walls in a 
two-dimensional antiferromagnet. This picture naturally leads to a spin-only 
description, since charge and spin are spatially segregated. As Castro Neto and
Hone \cite{CNH} pointed out, the only influence of the localized holes on the
spin system is to induce a weaker exchange-interaction across the stripes as 
compared to those in the magnetic domains. The system is therefore 
described by a model of coupled
Heisenberg spin-ladders, which is in the universality class of the QNLS. 

In addition, there are the options that the stripe phase is either bond
ordered, as suggested by exact diagonalization studies on the t-J model
\cite{White}, or site ordered,
as suggested by mean field calculations \cite{Jan}. In spin-only language 
this would correspond 
with coupled two-leg and three-leg spin ladders for bond- and site order, respectively.
We have performed an extensive analysis of the magnetic properties of these coupled ladder
models, finding that the three-leg ladder model can be excluded on basis of the available
experimental information. If the stripes are bond-ordered, the physics of coupled two-leg
ladders could be responsible for the quantum magnetism of the stripe phase, and a strategy
is pointed out to further investigate these matters by experiment. 
\\
\\
Before we turn to the microscopy of the coupled-ladder model, let us first discuss the 
long-wavelength behavior as follows from the QNLS \cite{CHN}. 
The QNLS describes the transversal 
spin-fluctuations of a system with well-formed local magnetic moments. 
It is characterized by a single coupling constant
($g$) which measures the strength of the quantum fluctuations. This
constant depends on the microscopic details. The behavior of the
model as a function of $g$ and temperature is shown in figure 1.

At zero temperature, long-range order is found for $g$ smaller than its
critical value, while the N\'eel state gets disordered at long distances for $g>g_c$. At 
finite temperatures, four distinct regimes are found, separated by cross-over lines.
For small $g$, the system is in the renormalized classical (RC) region, where the correlation
length diverges exponentially as the temperature is lowered towards the long-range
ordered phase at $T=0$. At large coupling, a quantum disordered (QD) region is found,
where the system has acquired a spin gap, while the correlation-length becomes 
independent of temperature. The above two regimes are separated by a quantum critical
region. Here, the system behaves as if it is at its zero-temperature
critical point on length-scales smaller than the inverse temperature (in appropriate 
units). The correlation length increases as $1/T$ upon cooling. More generally, the
only relevant energy scale in the system is temperature, which
gives rise to the so-called $\omega/T$-scaling (energy scales as temperature). 
This region terminates at a cut-off temperature, where the correlation 
length becomes of order of one lattice-spacing.  Non-universal effects become 
important and the continuum QNLS-description is no longer applicable.

Experiments on stripe-ordered systems are consistent with a QNLS-description where
$g$ is close to $g_c$. A first indication is found in measurements of the
magnetic ordering temperature. 
Due to small inter-plane interactions and spin-anisotropies, long-range magnetic order 
sets in at a
finite temperature, when the 2d correlation length reaches a large critical value.
In the stripe phase, this occurs at a much lower temperature than in the undoped 
antiferromagnet \cite{Tranq}. Since the scale on the vertical axis in fig. 1, which is set by the exchange coupling $J$, is of the same order as for zero doping 
\cite{Hayden}, this suggests a shift of $g$ to a larger value. Further indications
come from the work of 
Kataev {\em et. al.}, who performed ESR measurements on $Eu$ doped
$La_{2-x}Sr_x Cu O_4$, using a $Gd$ spin probe \cite{Kataev}. They find an exponential increase of the spin-lattice relaxation
time $1/T_1$ upon cooling below the charge-ordering temperature $T_{\rm co}=70 K$, 
signalling renormalized classical behavior. The correlation length near $T_{\rm co}$
is of order 10 lattice constants. This is larger than the stripe separation, suggesting that
a continuum QNLS description is valid, but still much smaller than the correlation length in
the undoped system at the same temperature. Again, this suggests that the ordered
stripe system is like the undoped antiferromagnet, but with considerably more 
quantum-fluctuations. For a material which has ordered stripes to high temperatures,
we therefore expect to get the behavior indicated by the left grey line in 
fig. 1.
\\
\\
The question we address is whether our simple spin-only model
of the ordered static stripe phase can account for
the observed quantum fluctuations in the spin sector.
This question, being related to {\it microscopy} of
stripe ordering and thus to non-universal properties 
of the system, cannot be answered purely on the grounds
of continuum field theory. We therefore resort to numerical
simulations of the coupled 2- and 3-leg spin ladder system
\cite{Jacub}. The highly efficient loop algorithm for
the Quantum Monte Carlo method \cite{QMC} allows us
to find the parameter dependences of the correlation length in
systems containing up to $1.6\times 10^4$ sites and 
temperatures as low as 0.03 $J$.  We extract the cross-over lines separating the different
regions of the QNLS diagram from the temperature dependence
of the correlation length. Our results are
summarized in fig.2. We denote the
entrance to the
renormalized classical regime by $T^*$, $T'$ marks the 
onset of Quantum Disordered behavior, and the "cut-off" $T_0$ is given by the temperature above which the spin dynamics becomes that of independent
single ladders. We also fit our results to the different
cross-over scales calculated form an anisotropic
QNLS model \cite{Jacub,Coen}, using 
\begin{figure}[th]
\epsfxsize= \hsize
\hspace{-.04 \hsize}
\epsffile{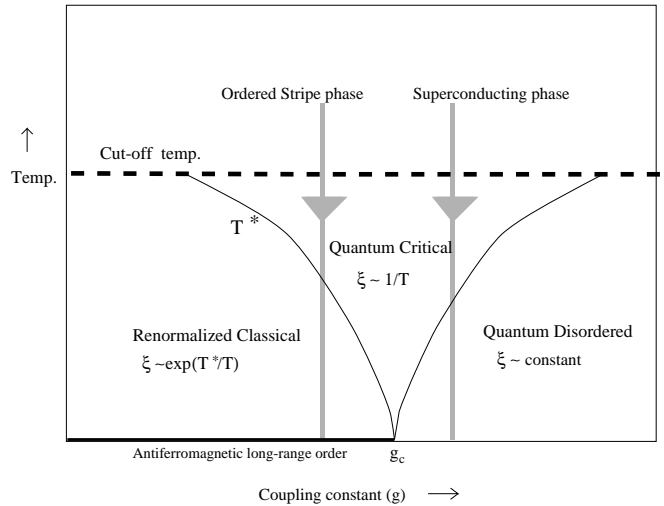}
\vspace{.2cm}
\caption{Cross-over diagram of the QNLS. The grey
lines indicate the measurement-trajectories as the system is
cooled towards the ordered 
stripe- and the superconducting phase.}
\label{fig1}
\end{figure}

\noindent
only the bare coupling 
constant as a fitting parameter.

The isolated 3-leg spin-ladder has a Luttinger liquid
ground state, characterized by algebraic decay of the correlations.
It is therefore expected that any nonzero inter-ladder 
coupling  $\alpha J$ ($J$ is the coupling in the ladder) suffices to
establish 2+1D antiferromagnetic order \cite{Afflek}. The
ground state is then in the renormalized classical regime
and the crossover scale $T^*$ acquires a finite value.
Our numerical data (Fig.2a) indeed show such a behavior.
One interesting observation concerns a non-universal
(in the sense of continuum field theory) feature:
$T^*$ coincides with the crossover $T^0$ for any anisotropy $\alpha$.
The classical behavior sets in at the very moment the 
system discovers that it is (2+1) dimensional. The 
experimental prediction for such a system would be
that no QC behavior could be observed, because
the region above RC is always governed by lattice cut-off
physics. 

This last conclusion turns out to be remarkably different
for the 2-leg coupled case. Since the isolated 2-leg ladder
possesses an energy gap, the ladder-to-ladder interaction
has to overcome this energy scale in order to get a (2+1)D 
behavior. With our numerical simulations (Fig.2b) we find
indeed that the quantum order-disorder transition occurs
at a finite and quite large value of anisotropy:
$\alpha = 0.30(2)$. In the vicinity of this anisotropy
scale the crossovers $T^0$ and $T^*$ (also $T'$)
separate and {\em a large QC regime opens up}.
The results may be quantitatively reproduced with an appropriate
\cite{Jacub,Coen} anisotropic QNLS model (see
Fig.2b).

We arrive at the following conclusions: in order to find a strong influence
of quantum fluctuations, the ratio of inter- to intra-ladder 
coupling has to be unrealistically close to zero for the case of site-ordered
stripes (3-leg ladders). Hence, the coupled
ladder model predicts {\em bond-ordered} stripes in
the cuprates, with $\alpha$ close to its critical value of $0.3$.
This gives for the anisotropy in the spin-wave velocities
parallel/perpendicular to the stripes
\[
\frac{c_{\perp}}{c_{\parallel}}= \sqrt{\frac{2\alpha_c}{1+\alpha_c}} \approx 0.7 \, .
\]

The results presented here provide a lowest-order description of the static
stripe state. It may be possible to test these results experimentally. The question of
site- or bond-ordered stripes can be addressed by NMR. This has already been done
for stripes in the Nickelate system $LaNiO_{4+\delta}$, which were found 
to be site-ordered \cite{Brom}. Neutron-scattering measurements of the spin-wave velocities 
along and across the stripes could establish the magnitude of the spatial anisotropy in the spin-spin
interactions. This would tell us how much of the quantum fluctuations in the spin-sector
can be attributed purely to this anisotropy.

\begin{figure}[h]
\vspace{-2.3cm}
\epsfxsize=0.9 \hsize
\epsffile{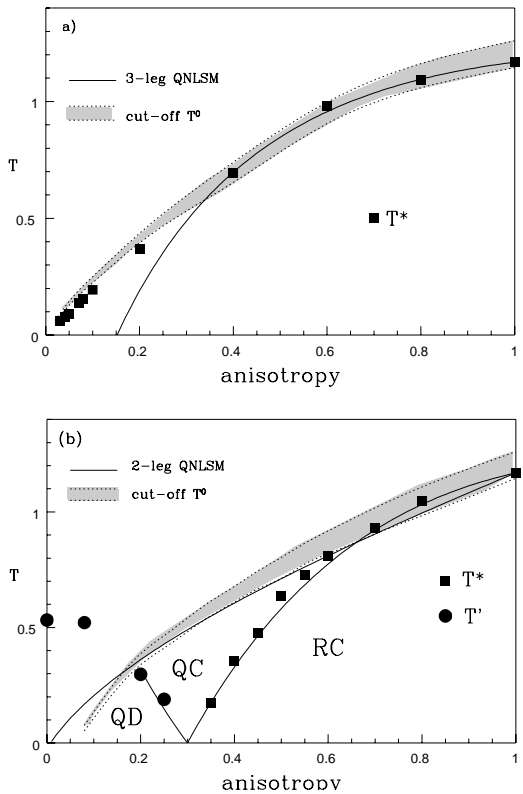}
\vspace{.2cm}
\caption{Temperature crossovers for a) coupled 3-leg
ladders b) coupled 2-leg ladders as a function of anisotropy.
Solid lines represent the result of fitting an analytic theory.}
\end{figure}

\acknowledgements{We acknowledge O. Y. Osman for usefull discussions and for his
contribution to the QMC calculations.}

\vspace{3cm}
$\,$

\end{multicols}
\end{document}